\documentclass[conference]{IEEEtran}
\IEEEoverridecommandlockouts
\usepackage{cite}
\usepackage{amsmath,amssymb,amsfonts}
\usepackage{acronym}
\usepackage{algorithmic}
\usepackage{graphicx}
\usepackage{textcomp}
\usepackage{xcolor}
\usepackage{url}
\def\BibTeX{{\rm B\kern-.05em{\sc i\kern-.025em b}\kern-.08em
    T\kern-.1667em\lower.7ex\hbox{E}\kern-.125emX}}
\usepackage[numbers]{natbib}
\usepackage{tikz}
\usetikzlibrary{decorations.text}
\usetikzlibrary{shapes}
\usepackage{pgfplots}
\usepackage{subfig}
\pgfplotsset{compat=1.18}
\begin{document}

\title{Realistic Neutral Atom Image Simulation
\thanks{This work was funded by the German Federal Ministry of Education and Research (BMBF) under the funding program \textit{Quantum Technologies - From Basic Research to Market} under contract numbers 13N16070, 13N16076, and 13N16087.}
}

\author{\IEEEauthorblockN{Jonas Winklmann}
\IEEEauthorblockA{\textit{Chair of Computer Architecture and Parallel Systems}\\
\textit{Technical University of Munich}\\
Munich, Germany\\
jonas.winklmann@tum.de}
\and
\IEEEauthorblockN{Dimitrios Tsevas}
\IEEEauthorblockA{\textit{Quantum Many-Body Systems Division} \\
\textit{Max Planck Institute of Quantum Optics}\\
Garching, Germany\\
dimitrios.tsevas@mpq.mpg.de}
\and
\IEEEauthorblockN{Martin Schulz}
\IEEEauthorblockA{\textit{Chair of Computer Architecture and Parallel Systems}\\
\textit{Technical University of Munich}\\
Munich, Germany\\
schulzm@in.tum.de}
}

\acrodef{EMCCD}[EMCCD]{electron-multiplying charge-coupled device}
\acrodef{CMOS}[CMOS]{complementary metal-oxide-semiconductor}
\acrodef{qCMOS}[qCMOS]{quantitative \ac{CMOS}}
\acrodef{MPQ}[MPQ]{Max Planck Institute for Quantum Optics}
\acrodef{sCIC}[sCIC]{serial clock-induced charge}
\acrodef{CIC}[CIC]{clock-induced charge}
\acrodef{SNR}[SNR]{signal-to-noise ratio}
\acrodef{EM}[EM]{electron-multiplying}
\acrodef{MTF}[MTF]{modulation transfer function}
\acrodef{PSF}[PSF]{point-spread function}
\acrodef{OTF}[OTF]{optical transfer function}
\acrodef{CDF}[CDF]{cumulative distribution function}
\acrodef{PDF}[PDF]{probability density function}
\acrodef{GCC}[GCC]{GNU Compiler Collection}
\acrodef{ROI}[ROI]{region of interest}
\acrodef{ADC}[ADC]{analog-to-digital converter}
\acrodef{MQV}[MQV]{Munich Quantum Valley}

\IEEEpubid{\vspace{1cm}\begin{minipage}{\textwidth}\ \\[24pt] \centering
  979-8-3503-4323-6/23/\$31.oo \copyright 2023 IEEE. Personal use of this material is permitted. Permission from IEEE must be obtained for all other uses, in any current or future media, including reprinting/republishing this material for advertising or promotional purposes, creating new collective works, for resale or redistribution to servers or lists, or reuse of any copyrighted component of this work in other works. DOI 10.1109/QCE57702.2023.00153
\end{minipage}}

\maketitle

\begin{abstract}
Neutral atom quantum computers require accurate single atom detection for the preparation and readout of their qubits. This is usually done using fluorescence imaging. The occupancy of an atom site in these images is often somewhat ambiguous due to the stochastic nature of the imaging process. Further, the lack of ground truth makes it difficult to rate the accuracy of reconstruction algorithms.

We introduce a bottom-up simulator that is capable of generating sample images of neutral atom experiments from a description of the actual state in the simulated system. Possible use cases include the creation of exemplary images for demonstration purposes, fast training iterations for deconvolution algorithms, and generation of labeled data for machine-learning-based atom detection approaches.
The implementation is available through our GitHub as a C library or wrapped Python package.

We show the modeled effects and implementation of the simulations at different stages of the imaging process. Not all real-world phenomena can be reproduced perfectly. The main discrepancies are that the simulator allows for only one characterization of optical aberrations across the whole image, supports only discrete atom locations, and does not model all effects of \ac{CMOS} cameras perfectly.

Nevertheless, our experiments show that the generated images closely match real-world pictures to the point that they are practically indistinguishable and can be used as labeled data for training the next generation of detection algorithms.
\end{abstract}


\section{Introduction}
\begin{figure*}
\centering
  \subfloat[]{\includegraphics[width=0.3\linewidth]{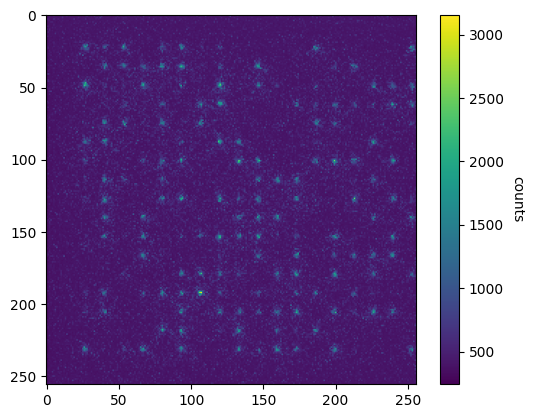}\label{figures:exampleImageReal}}
  \hspace{1cm}
  \subfloat[]{\includegraphics[width=0.3\linewidth]{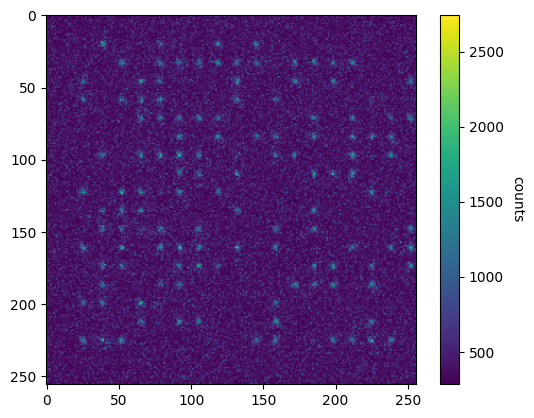}\label{figures:exampleImageSimulated}}
\caption{(a) Exemplary image of an array of atoms taken by an Andor iXon Ultra 897 (b) Simulated image using settings calibrated on real image}\label{figures:exampleImage}
\end{figure*}
In neutral atom quantum computers, typically, the way of extracting information from the system is through fluorescence imaging \cite{Morgado:1}. An \ac{EMCCD} or \acf{CMOS} camera takes an image of the array of atoms and a reconstruction algorithm then analyzes the image to detect the occupancy of each potential atom site. An example of such an image can be seen in Fig.~\ref{figures:exampleImageReal}. However, no ground truth exists for all images taken on real systems and the only means of knowing about the state of the system is to trust the reconstruction algorithm.

%this means that taking an image and analyzing it is the only way of knowing anything about the system's state and no ground truth exists.

When imaging the experiment, Poissonian-distributed photoelectrons from atomic fluorescence compete with stochastically distributed electrons from camera noise of various origins. In a performance-tuned quantum computer, the qubit readout time is minimized and so only few fluorescence photons per atom can be collected. The aforementioned competition between photoelectrons and noise electrons radically reduces our confidence in determining whether a site contains an atom or not, and with that our confidence in the final image analysis.

This problem is further intensified by the potential loss of atoms during the imaging process if it is not tuned perfectly. The sites of these lost atoms look like normal occupied sites with all the characteristics of an atom, except their effective exposure time may assume any value between ``no time at all'' and ``the full exposure time''. There may, therefore, be a continuous transition between the brightness of unoccupied and occupied sites. At the end, though, the only noteworthy metric is whether a site is occupied after taking the image, which is a binary decision. While this can be enhanced by taking consecutive images, these again lack certainty. Sites that are inherently dimmer or appear dimmer by chance might be omitted from testing on real data, despite being the most interesting ones, since assumptions about their occupancy are probabilistic.

Optimized detection algorithms are required to increase detection fidelity under these challenging experimental conditions, but their design and training requires labeled data with absolute knowledge of atomic presence or absence to evaluate their performance accurately, which we cannot gain from real-world experiments.
\IEEEpubidadjcol
In this work, we, therefore, propose a different way of generating artificial images with realistic properties directly from the corresponding ground truth. Starting from a given array of atom locations and states, our approach constructs a picture from the ground up simulating the imaging process. Since the process of image formation, considering influences from both photoelectrons and noise electrons, is well understood, we can produce atomic fluorescence pictures that are practically indistinguishable from real ones. As a second benefit, the simulated pictures can also be calculated much faster than real ones can be obtained experimentally, allowing for rapid development and training on top of our simulator. Fig.~\ref{figures:exampleImageSimulated} shows such a simulated image.

Since the simulation output comes with absolute truth of occupancy, from which it was created, it can be used to benchmark detection algorithms accurately by comparing the reconstructed state to the original one. However, one needs to have trust in the simulator’s capabilities of producing accurate images. All noise sources and probability distribution have to be modeled and sampled accurately in order to rate an algorithm's performance using the generator's data.

After the simulation's implementation, we show its output's differences and similarities to actual images. Processes that have to be modeled accurately include:
\begin{enumerate}
    \item The probability distribution of filling the array with atoms and the chance and timing of atoms being lost during imaging.
    \item The properties of the optical system and any imperfections or aberrations that a physical camera may have to deal with.
    \item The internal workings of different cameras from the physical pixel to the digital output.
\end{enumerate}

The real data on which we build our approach and which we use for evaluation stems from experiments at the Strontium Rydberg Lab at \ac{MPQ}. Section~\ref{chapters:process} gives a short explanation of a nearly equivalent setup. It is to be noted, though, that some of the data comes from very early stages of the overarching project or was acquired through intentional parameter choices that may cause images to appear subpar to state-of-the-art imaging techniques. The development of the generator sometimes required data that is usually not available when optimizing the imaging system. There are, for example, not enough atoms lost in imaging to make statements about their brightness distribution if one does not actively produce them. That is to say none of the data used in this paper should reflect badly on \ac{MPQ}.

This generator is produced open-source and can help with getting sample images as well as developing accurate reconstruction algorithms. In order to allow for a short execution time as well as ease of use, the core functionality is implemented in C, while a Python wrapper improves accessibility.

Our contributions include:
\begin{itemize}
    \item We investigate the noise sources involved in imaging an array of atoms.
    \item We develop algorithms that efficiently model these effects and use them to implement a comprehensive and accessible tool that can generate sample images.
    \item We show how to analyze real image datasets in order to extract parameters for tuning the algorithm's output.
    \item We evaluate the accuracy of the simulator by investigating its ability to reproduce certain image characteristics and underlying stochastics.
\end{itemize}

Overall, the simulator works as intended and delivers practically indistinguishable images, compared to real images, for the listed use cases. While there are minor inconsistencies like unexplained artifacts of \ac{CMOS} cameras, only globally configurable optical aberrations, and discrete atom positioning, all major effects are modeled leading to realistic images. 

%Real images may exhibit a range of parameters throughout a single image that cannot be reproduced within one simulated image. In that situation, the user is advised to generate a dataset using changing parameters.
\section{Imaging Process}\label{chapters:process}
As part of a larger initiative to develop a neutral atom quantum computer, we use a 2D rectangular grid of strontium atoms that act as qubits. Since loading the atoms into position is inherently probabilistic, we must first establish which sites are initially occupied after the loading process, so that in a second step we can create a completely filled array by moving specific atoms from occupied locations around. Both this initial occupancy detection, as well as any qubit readout require imaging the array.
\begin{figure}
\centering
    \begin{tikzpicture}[>=latex]
    \node (image) at (0,0) {\includegraphics[width=\linewidth]{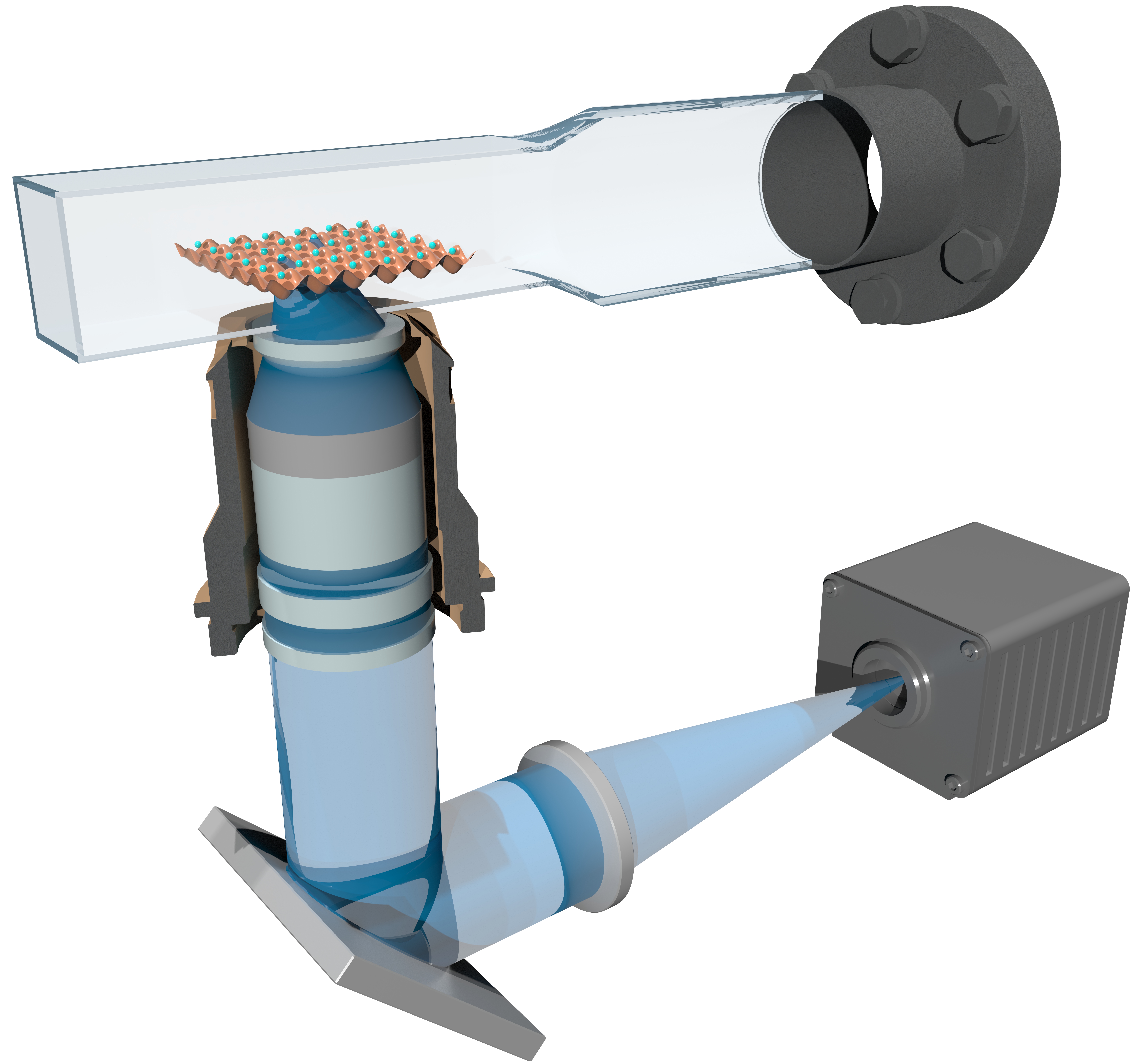}};
    \node (atomText) at (-3,3.8) [rectangle callout,draw=gray!80,fill=gray!20,align=center,callout absolute pointer={(-2.4,2.32)}] {Strontium\\atoms};
    \node (glassCellText) at (1.2,4) [rectangle callout,draw=gray!80,fill=gray!20,align=center,callout absolute pointer={(1.2,3.35)}] {Glass\\cell};
    \node (trapText) at (-0.7,4) [rectangle callout,draw=gray!80,fill=gray!20,align=center,callout absolute pointer={(-0.95,2)}] {Optical\\dipole traps};
    \node (cameraText) at (3,0.8) [rectangle callout,draw=gray!80,fill=gray!20,align=center,callout absolute pointer={(3,0.06)}] {CMOS\\camera};
    \node (lensText) at (2.5,-3) [rectangle callout,draw=gray!80,fill=gray!20,align=center,callout absolute pointer={(0.45,-2.7)}] {Converging\\lens};
    \node (objectiveText) at (1,0.5) [rectangle callout,draw=gray!80,fill=gray!20,align=center,callout absolute pointer={(-0.85,0.5)}] {Microscope\\objective};
    \end{tikzpicture}
    \caption{Schematic illustration of the planned experimental setup in the \ac{MQV} lab at \ac{MPQ}. Only the elements that are relevant for fluorescence imaging are shown. The illustration omits the global laser beam that is resonant to an electronic imaging transition in the strontium atoms and that excites the fluorescence light. Illustration by Dr. Andrea Alberti, \ac{MQV} team at \ac{MPQ}. Annotated by Jonas Winklmann.} 
    \label{figures:setupSchematic}
\end{figure}
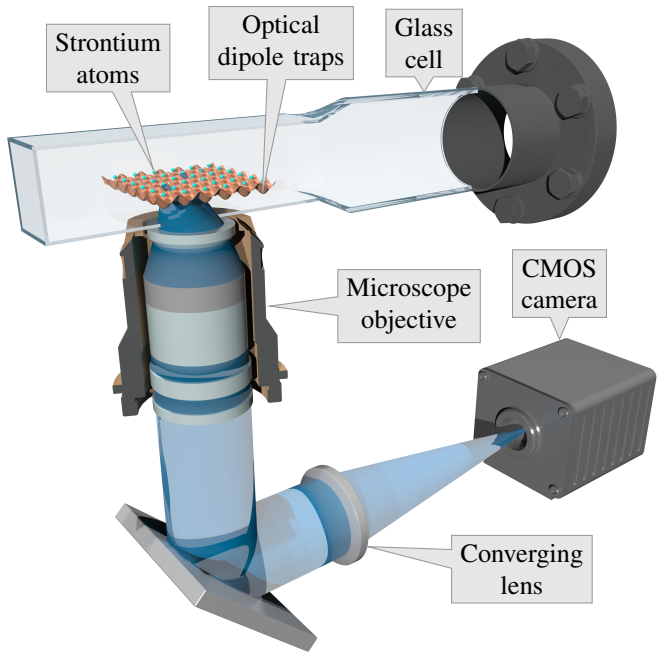

Fig.~\ref{figures:setupSchematic} illustrates the setup at \ac{MPQ}'s \ac{MQV} team, which resembles the one that we used 
%. All image data used in developing the simulator was 
to acquire all data for our developments.
%through a equivalent or similar setup. 
Single strontium atoms are spatially confined at distinct and discrete locations within a single plane by an array of optical dipole traps inside a glass cell under vacuum. An infinity- and aberration-corrected microscope objective and a converging lens are placed in a 4-f-configuration with respect to the atomic plane on the one hand and with respect to the sensor of a \ac{qCMOS} camera on the other. Each atom can be approximated as an isotropic point source of fluorescence light. The illustration shows how the fluorescence light of two arbitrary atoms is collimated by the objective and then focused onto the camera by the lens. The objective and the lens together can be approximated as a paraxial imaging system, which we describe with Fourier optics methods in the simulation.
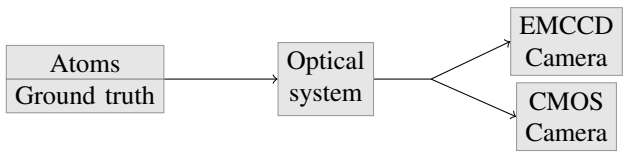
\begin{figure}
\centering
    \begin{tikzpicture}
        %Nodes
        \node (states) at (0, 0) [draw=gray!80,fill=gray!20,align=center] {Atoms\\Ground truth};
        \node (optics) at (3.2, 0) [draw=gray!80,fill=gray!20,align=center] {Optical\\system};
        \node (cmos) at (6.4, -0.5) [draw=gray!80,fill=gray!20,align=center] {CMOS\\Camera};
        \node (emccd) at (6.4, 0.5) [draw=gray!80,fill=gray!20,align=center] {EMCCD\\Camera};
        %Lines
        \draw[->] (states.east) to (optics.west);
        \draw[-] (optics.east) to (4.6, 0);
        \draw[-,draw=gray!100] (states.west) to (states.east);
        \draw[->] (4.6, 0) to (cmos.west);
        \draw[->] (4.6, 0) to (emccd.west);
    \end{tikzpicture}
    \caption{Sequence of steps during imaging}
    \label{figures:imagingSteps}
\end{figure}

Fig.~\ref{figures:imagingSteps} shows the sequence of logical steps that are involved in the formation of images. To begin with, there are the atoms themselves. Each atom location has a certain probability of being occupied. Whether or not a given trap is filled can be determined via fluorescence imaging \cite{Schlosser:1}, which allows for resolving single atoms due to its high \ac{SNR} ratio \cite{Morgado:1}. If the imaging technique is not yet developed to perfection, there may be some atoms that escape from their traps while we take the picture. The optical setup then has to focus the fluorescence light onto the camera. The positioning of all relevant atoms within the camera's field of view, their alignment with image axes, magnification, and optical aberrations are factors that have to be considered here.
%\begin{figure}
%  \subfloat[]{\includegraphics[width=0.48\linewidth]{images/ixon-ultra-897-product.png}}
%  \hfill
%  \subfloat[]{\includegraphics[width=0.48\linewidth]{images/s_c15550-20up_pp,0.jpg.thumb.252.252.jpg}}
%\caption{(a) Andor iXon Ultra 897 \cite{Andor:1} (b) ORCA-Quest \cite{Hamamatsu:3}} \label{figures:cameras}
%\end{figure}

Once the light has reached the camera, some of the photons hitting the sensor generate a charge. These charges are called primary photoelectrons and represent the discretized light intensity. Starting from this number of charges, internal camera effects take place. \ac{EMCCD} cameras, for example, multiply the number of photoelectrons, albeit at the cost of increased noise. At the end, the value is read out and written into the final image. In this project, the cameras used to take images of the setup are the \ac{EMCCD} camera Andor iXon Ultra 897 and the \ac{qCMOS} camera ORCA-Quest.
\section{Modeling}
Having discussed the process of taking images of an array of atoms, next we can explain how we model each stage in our simulator to capture the needed physical effects and to which degree the output is analogous to real images being taken.

\subsection{Stage 1: Modeling the Atom Experiment}\label{chapters:experiment}
In the initial stage of our algorithm we first determine the user-specified configuration, from which an image will be created. This includes the configured atom locations, the filling ratio, and other parameters to determine actual occupancy. The user is further given the choice of whether they want to decide the occupancy of each site themselves or let the program fill each location randomly. Depending on the use case, these locations may be either a list of image coordinates or information describing the grid of atoms. 

Based on the data, we then prepare an initial 2D array covering the size of the camera's resolution and prepare it by setting all values to 0 and adding 1 for each atom at its appropriate locations in camera coordinates. These values are then reduced for some sites to simulate the effects of atom loss. For each occupied location, a random value determines whether the atom is lost and, if so, the pixel value is decreased. Any loss of atoms is independent of the time and, therefore, modeled via exponential decay. 

At the end, the final information is stored separately and can later be used as ground truth for the generated image.

%This value between 0 and 1 is also reported back to the user as the advertised ground truth.

\subsection{Stage 2: Modeling the Optical System}\label{chapters:optics}
We now simulate the image response to the atoms, both due to the wave properties of light, as well as defects of the optical system. To do so, the array from before is convoluted with the \ac{PSF} of the optical system. For that, the \ac{MTF}, i.e. the magnitude of the Fourier transform of the \ac{PSF}, is calculated. First, the complex pupil function is constructed. Outside of its radius, this pupil is set to 0 while the values inside the radius hold the phase values, which are derived from Zernike polynomials. This allows for accurate modeling of optical aberrations and defects of the setup. The coefficients of the individual Zernike terms can be acquired from fitting real data. Applying a Fourier transform to this pupil and squaring the absolute values of the result yields the \ac{PSF}. Calculating the Fourier transform of this \ac{PSF} results in the \ac{OTF}~\cite{Goodman:1}, of which we again use the absolute values. These are then normalized and the zero-frequency component is shifted to the edges, yielding the final \ac{MTF}.

Next, the array from Section~\ref{chapters:experiment} is Fourier-transformed and multiplied with the \ac{MTF}. We now apply an inverse Fourier transform to this and take the absolute values. Afterwards, we scale the pixel values in order for the sum of the pixel values to match the sum of the original image, yielding the final 2D photon distribution from the imaged atoms. This is then multiplied with the expected number of photons per atom. The final result represents the expected number of photons to hit each pixel. With that, we have simulated all effects that influence the image before the light actually reaches the camera.

\subsection{Stage 3: Modeling Camera Effects}\label{chapters:cameraProcesses}
In this stage, we introduce camera effects that can influence the final image. Starting with the more complicated camera type, we explain the noise sources of \ac{EMCCD} cameras first. In addition to the expected photons, which we calculate for each pixel in Section~\ref{chapters:optics}, other electron sources may exist. These charges can stem from background illumination, dark current, and \acp{CIC}. To capture these effects we rely on the camera's datasheet, which provides values for the dark current rate and the rate of \acp{CIC}, while fitting pixel histograms of real pictures yields the rate of stray light. We then use the final sum of all charges in sampling the discrete number of resulting photoelectrons, which follow a Poisson distribution~\cite{Seitz:1}. The noise that is introduced due to the Poissonian nature of the photon count is called shot noise.

Another effect that needs to be taken into account is the process of \ac{EM}. Through avalanche multiplication, the number of primary photoelectons can be multiplied in so-called gain registers, resulting in a linear gain on average. With this, however, also the error increases. Additionally, it is possible for a secondary charge to emerge spontaneously without a primary photoelectron. This is called a \ac{sCIC}. At last, the value is read out. For this to be realistic in our simulation, we apply an offset (bias clamp) and a scale factor (preamp gain). The resulting value serves as the mean of a Gaussian distribution that is sampled to acquire the actual pixel value.

A \ac{CMOS} camera simplifies this process greatly. Through high quantum efficiency, low dark current rates, and non-existent \acp{CIC}, modern \ac{CMOS} cameras are able to provide images with even better \acp{SNR} than \acp{EMCCD}, without introducing noise through electron multiplication. Since there are no \acp{CIC}, \acp{sCIC} or \ac{EM} gain noises, only shot noise, dark current, and readout noise remain~\cite{Hamamatsu:2}. The main disadvantage of this technology is that each pixel behaves slightly differently. However, these drawbacks can easily be remedied through calibration since the pixel-specific parameters remain constant for each pixel. In the simulation, we, therefore, assign each pixel random values for offset, gain, and dark current rate.

\begin{figure}
\centering
  \includegraphics[width=0.72\linewidth]{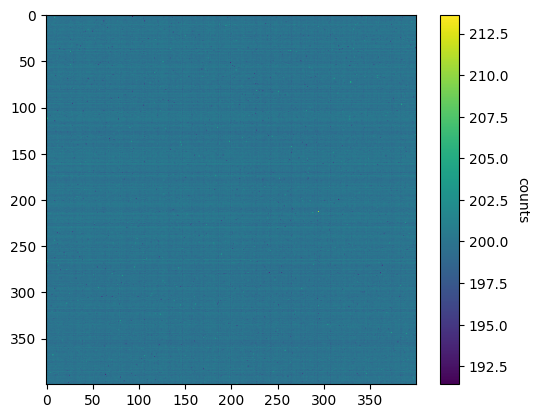}
    \caption{Average over 417 images taken by an ORCA-Quest with a closed shutter and 10ms exposure time} 
    \label{figures:CMOSClosedShutter}
\end{figure}

Complicating matters further, there is some structure to this noise. For example, pixels within a row and column often exhibit similar values. In Fig.~\ref{figures:CMOSClosedShutter}, these rows and columns are clearly visible. While this makes it more complicated, values that take this structure into account can still be assigned to each individual pixel and the processes that lead to these discrepancies are well understood.

\begin{figure}
\centering
  \includegraphics[width=0.8\linewidth]{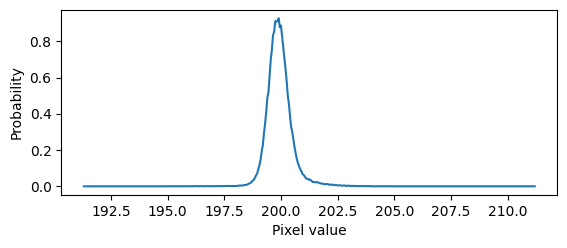}
    \caption{Histogram over mean values of each pixel for a dataset of 438 images without atoms with a binning width of 0.05} 
    \label{figures:CMOSPixelMeans}
\end{figure}

The dark current rate follows a Gamma distribution~\cite{Baer:1}. The offset value can be estimated by plotting the mean value for each pixel over a dataset with closed shutter and a low exposure time of 1ms. This distribution can be seen in Fig.~\ref{figures:CMOSPixelMeans}. Readout noise distributions for different camera models are shown in~\citeauthor{Hamamatsu:2}'s white paper and are modeled by~\citeauthor{Fowler:1}~\cite{Hamamatsu:2,Fowler:1}. While this model seems to be quite accurate, it require 16 parameters to tune. Since it is not practical to have the user finely tune this many inputs, we use a simplified model similar to~\citeauthor{Gow:1}'s solution. Older camera models often had a significant number of pixels with outstandingly large readout noise, which led~\citeauthor{Gow:1} to merge two probability distributions for modeling noise. As these tails of the readout noise distribution have shortened considerably for modern \ac{CMOS} cameras, we only use the Gumbel distribution for sampling column and flicker noise \cite{Hamamatsu:2,Gow:1}. Row noise is modeled using a zero-mean Gaussian distributions. Since thermal noise and other unstructured Gaussian noise sources are indiscernible, they are combined into a single Gaussian noise in order to keep the parameter count manageable.

Overall, we capture all relevant physical processes in the above stages, which then allows us to create realistic images that are practically indistinguishable from real images. 
\section{Sampling of Probability Distributions}
The imaging process involves many different stochastic processes that require accurate sampling in order for the simulation to be trustworthy. We implement the required Poisson and Gaussian distributions using \citeauthor{Knuth:1}'s method for counting inter-arrival times~\cite{Knuth:1} and the Box-Muller transform~\cite{Box:1}, respectively. The Gamma distribution, which we use to represent the distribution of dark current rates in \ac{CMOS} cameras, is sampled using \citeauthor{Marsaglia:1}'s method \cite{Marsaglia:1}. 

%Neither of these need to be proven since they are well established.

We sample the remaining probability distributions using the inverse cumulative method. The inverse of the \ac{CDF} is calculated or estimated and a random variable between 0 and 1 is used as the parameter, yielding a random variable that follows the original \ac{PDF} \cite{Devroye:1}.

\subsection{Atoms Lost During Imaging}\label{chapters:atomLossDuringImagingDist}

As mentioned in Section~\ref{chapters:experiment}, the atoms that are lost during imaging can be modeled as exponential decay. Since the number of atoms decreases over time, less atoms are lost late in the imaging process. The time when an atom was lost can be sampled as follows.

The time t that will be used in the following equations represents the fraction of the imaging process that the atom survived, meaning 0 corresponds to the atom being lost right at the beginning and 1 right at the end. The probability of it being lost at a given moment is relative to the number of atoms remaining. The \ac{PDF}, normalized so that its integral between 0 and 1 equals 1, is given as
\begin{equation}
    \begin{aligned}
PDF_p(t)=\frac{p^t}{\int_0^1p^tdt} = \frac{p^t\cdot ln(p)}{p-1}
    \end{aligned}
\end{equation}
with p being the probability that a given atom stays in its trap during the whole imaging process. We call this the survival probability. As part of the inverse cumulative method, the inverse \ac{CDF} is calculated and a random variable, uniformly distributed between 0 and 1, is inserted. The functions are given by
\begin{equation}
    \begin{aligned}
CDF_p(t)=\int_0^t\frac{p^t\cdot ln(p)}{p-1}dt = \frac{p^t-1}{p-1}
    \end{aligned}
\end{equation}
\begin{equation}
    \begin{aligned}
CDF_p^{-1}(r) = log_p((p-1)\cdot r+1)
    \end{aligned}
\end{equation}
with r being the uniformly distributed random variable. One can change the logarithm's base if an unknown one is undesirable. For high survival probabilities, r is increasingly being scaled down. The mapping from random variable to sampled time, therefore, becomes nearly linear. This makes sense since the decay is so slow that there are nearly as many atoms left at the end as there are at the beginning. For lower survival probabilities, this mapping is skewed towards earlier moments in time.
\subsection{\ac{EM} Gain}
The \ac{EM} gain process produces an increased number of secondary photoelectrons for a given number of incoming primary ones. The probability for there being n secondary electrons given x primary ones and an \ac{EM} gain g is described by
\begin{equation}\label{eq:emGain}
    \begin{aligned}
P(n|x) = \frac{n^{x-1}\cdot e^{-\frac{n}{g}}}{g^x\cdot(x-1)!}
    \end{aligned}
\end{equation}
as provided in \cite{Alberti:1}. This function closely resembles a Poisson distribution, albeit a continuous interpretation.
\begin{equation}
    \begin{aligned}
CDF(n|x) = \int_{0}^{n} \frac{a^{x-1}\cdot e^{-\frac{a}{g}}}{g^x\cdot(x-1)!}da\\
= \frac{1}{g^x\cdot(x-1)!}\cdot \int_{0}^{n} a^{x-1}\cdot e^{-\frac{a}{g}}da\\
= \frac{1}{(x-1)!}\cdot (\Gamma(x,0)-\Gamma(x,\frac{n}{g}))\\
    \end{aligned}
\end{equation}

$\Gamma(a,b)$ denotes the incomplete gamma function ${\Gamma(a,b)=\int_{b}^{\infty}t^{s-1}\cdot e^{-t}dt}$. As a special case, $\Gamma(a,0)=\Gamma(a)$ with $\Gamma(a)$ being the complete gamma function, which is defined as $\Gamma(a)=(a-1)!$ for every positive integer a.
Since x represents the number of primary photoelectrons, it is, in fact, a positive integer.
\begin{equation}
    \begin{aligned}
CDF(n|x) = \frac{1}{(x-1)!}\cdot((x-1)! - \Gamma(x,\frac{n}{g}))\\
= 1 - \frac{\Gamma(x,\frac{n}{g})}{\Gamma(x)}
= 1 - Q(x,\frac{n}{g})
    \end{aligned}
\end{equation}
with $Q(a,b) = \frac{\Gamma(x,\frac{n}{g})}{\Gamma(x)}$ being the upper regularized gamma function. The number of secondary electrons can be sampled by solving $r = 1 - Q(x,\frac{n}{g})$ for $\frac{n}{g}$. The random variable r is uniformly distributed between 0 and 1. To get the inverse of the regularized gamma function, third-order Schröder iteration is used \cite{DiDonato:1}. These iteration steps take the form
\begin{equation}
    \begin{aligned}
t_{i+1} = t_i - \frac{f(t_i)}{f'(t_i)}\cdot (1+\frac{f(t_i)\cdot f''(t_i)}{2\cdot [f'(t_i)]^2})
    \end{aligned}
\end{equation}
as is shown by \citeauthor{Drakopoulos:1} \cite{Drakopoulos:1}. It is known that the average outcome for n should be the number of primary electrons multiplied by the \ac{EM} gain. The Schröder iteration calculates a sampled value for $\frac{n}{g}$, so its average result should be the number of primary electrons. Therefore, this value is chosen as the initial value $t_0$. At the end, the result will be multiplied by the \ac{EM} gain g to get the final number of electrons.

Since these iterations are designed to find the roots of a function, the sampling problem can be brought to the form $0 = 1 - r - Q(x,\frac{n}{g})$. As $r$ and $1 - r$ are equally distributed, this can be simplified to ${0=r-Q(x,\frac{n}{g})}$.\\
The components for the Schröder iteration are as follows.
\begin{equation}
    \begin{aligned}
f(t_i) = r - Q(x,t_i)\\
f'(t_i) = \frac{(t_i)^{x-1}\cdot e^{-t_i}}{\Gamma(x)}\\
f''(t_i) = \frac{(t_i)^{x-2}\cdot e^{-t_i}\cdot (x - t_i - 1)}{\Gamma(x)}
    \end{aligned}
\end{equation}

To minimize computation time, it is useful to precompute certain terms within each iteration step.
\begin{equation}
    \begin{aligned}
\frac{f''(t_i)}{f'(t_i)} = \frac{x-1}{t_i} - 1
    \end{aligned}
\end{equation}
\begin{equation}\label{eq:fOverDeriv}
    \begin{aligned}
\frac{f(t_i)}{f'(t_i)} = (t_i)^{1-x}\cdot e^{t_i}\cdot (r\cdot \Gamma(x) - \Gamma(x,t_i))
    \end{aligned}
\end{equation}

In this function, the upper incomplete gamma function $\Gamma(a,b)$ still appears. To solve it for positive integers a, the special case $\Gamma(1,b)=e^{-b}$ can be combined with the recurrence relation $\Gamma(a+1,b) = b\cdot \Gamma(a,b) + b^a \cdot e^{-b}$ to yield
\begin{equation}
    \begin{aligned}
\Gamma(a+1,b) = a!\cdot e^{-b}\cdot \sum_{k=0}^{a}\frac{b^k}{k!}
    \end{aligned}
\end{equation}
as is mentioned in \cite{Nist:1}. With this knowledge, (\ref{eq:fOverDeriv}) can be further simplified.
\begin{equation}
    \begin{aligned}
\frac{f(t_i)}{f'(t_i)} = (t_i)^{1-x} \cdot (x-1)!\cdot r\cdot e^{t_i} - \sum_{k=0}^{x-1}\frac{(x-1)!}{t_i^{x-k-1}\cdot k!}
    \end{aligned}
\end{equation}

The existence of factorials and high exponents makes overflows during code execution not unlikely. To prevent this, both the minuend and subtrahend of $\frac{f(t_i)}{f'(t_i)}$ are calculated iteratively. This way, factorials are never computed on their own.

Since both $\frac{f(t_i)}{f'(t_i)}$ and $\frac{f''(t_i)}{f'(t_i)}$ are now known, it is possible to compute t to any desired precision. However, as only integer numbers of secondary electrons make sense, it is not necessary to calculate very precise values. In the current implementation, the iteration runs as long as 
\begin{equation}
    \begin{aligned}
(t_{i+1}-t_i)^2 < \frac{1}{100\cdot g^2}
    \end{aligned}
\end{equation}

The division of the threshold by $g^2$ stems from the fact that the final value will be multiplied by g, resulting in an overall constant precision, independent of the \ac{EM} gain.

In many cases, e.g., for stray photons or \ac{sCIC} charges, there is exactly one primary photoelectron. For this special case, $Q(1,\frac{n}{g}) = e^{-\frac{n}{g}}$ and the value can directly be sampled as $-g\cdot ln(r)$.
\subsection{Gumbel Distribution}
The \ac{CDF} of the Gumbel distribution and its inverse are given by
\begin{equation}
    \begin{aligned}
CDF(x,\mu,\beta) = e^{-e^{-\frac{x-\mu}{\beta}}}
    \end{aligned}
\end{equation}
\begin{equation}
    \begin{aligned}
CDF^{-1}(r,\mu,\beta) = \mu - \beta\cdot ln(-ln(r))
    \end{aligned}
\end{equation}

If a zero-mean Gumbel distribution is required, then only the scale parameter $\beta$ may be configured. The location parameter is set to $\mu = -\beta\cdot\gamma$ with $\gamma$ denoting the Euler-Mascheroni constant.
\section{Fitting Parameter Values}
When using our developed simulation tool, one still has to provide parameters that are specifically chosen to allow for an accurate simulation of a particular experimental setup. This section explains the process behind the fitting of the most important parameters based on real data. In this case, images from an \ac{EMCCD} camera were used to calibrate each value.

\subsection{Atom Grid}
There are several parameters that are required to fully describe the imaged experiment. Along with the potential atom sites, characteristics include brightness, imaging survival probability, and filling ratio.
\begin{figure}
\centering
  \subfloat[]{\includegraphics[width=0.8\linewidth]{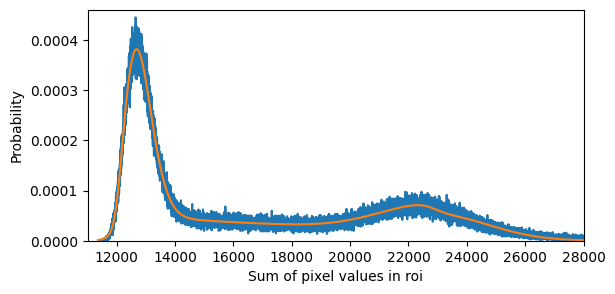}\label{figures:histogramFittingReal}}\\
  \subfloat[]{\includegraphics[width=0.8\linewidth]{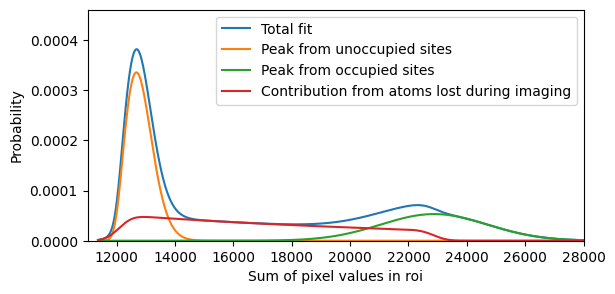}\label{figures:histogramFittingContributions}}
\caption{(a) Histogram over sum of pixel values in atom site \acp{ROI} (b) Fitted contributions to total distribution of sum of pixel values in atom site \acp{ROI}}\label{figures:histogramFitting}
\end{figure}

Looking at a set of real images, one can sum up the pixel values within a \acf{ROI} around each potential atom site. Fig.~\ref{figures:histogramFittingReal} shows the histogram for such \ac{ROI} sums for 2,839 images, taken using an \ac{EMCCD} camera, with 306 atom sites each, along with a fitted distribution. This distribution is the sum of the theoretical contributions of unoccupied sites, occupied sites, and ones where the atom was lost during imaging.

To set the radius of the \ac{ROI}, we chose that the innermost bright spot in the airy disk of a perfect optical response should be included completely. The radius of the first dark ring in focal plane coordinates is given by
\begin{equation}
    \begin{aligned}
r = 1.22\cdot\frac{\lambda\cdot z}{2\cdot r_a}
    \end{aligned}
\end{equation}
with $r_a$ being the radius of the aperture and $z$ the observation distance \cite{Goodman:1}.

Under the paraxial approximation, the infinity- and aberration-corrected microscope objective can be treated as a perfect thin lens with an effective focal length $z$. Combining ${\text{NA}=n\cdot sin\theta=n\cdot\frac{r_a}{\sqrt{r_a^2+z^2}}}$ with the index of refraction of the surrounding air $n\approx1$ and the inequality $z>r_a>0$ yields $\frac{r_a}{\sqrt2\cdot z}<\text{NA}<\frac{r_a}{z}$. Therefore,
\begin{equation}
    \begin{aligned}
1.22\cdot\frac{\lambda}{2\text{NA}} < r < 1.22\cdot\sqrt2\cdot\frac{\lambda}{2\text{NA}}
    \end{aligned}
\end{equation}
for wavelength $\lambda$ and numerical aperture NA. In our case, $0.433\mu m<r<0.613\mu m$. Multiplying by the magnification of our optical system of 156.25 and dividing by the physical size of a binned pixel results in a radius of between 2.12 and 2.99 pixels. Using this new information, the radius of the \ac{ROI} was chosen to be 3 pixels.

To analyze the individual distributions seen in Fig.~\ref{figures:histogramFittingContributions}, it is to be noted that each distribution is convoluted with a Gaussian due to readout noise. The readout distribution is given by
\begin{equation}
    \begin{aligned}
PDF_{readout}(x) = \frac{1}{\sigma\cdot\sqrt{2\pi}}\cdot e^{-\frac{1}{2}(\frac{x-\mu}{\sigma})^2}
    \end{aligned}
\end{equation}
for a standard deviation $\sigma$, which is given in the camera's datasheet, and the expected pixel value $\mu$. If all sites were unoccupied, the probability distribution would be given by
\begin{equation}
    \begin{aligned}
PDF_{occupied}(x) = PDF_{readout} \ast PDF_{Poisson,s,l}(x)
    \end{aligned}
\end{equation}
with $PDF_{Poisson,s,l}$ describing the Poisson distributed spurious charges from camera noise and background illumination after it is shifted and scaled along the x-axis to account for \ac{EM} gain and readout offset.

For occupied sites, the distribution is the same, only differing by the mean value of the Poisson distribution of charges. Since its mean is, however, quite high, it can be estimated as a Gaussian. Due to the convolution of two Gaussian distributions again being Gaussian, the whole distribution can be assumed to be Gaussian.

For atom locations where the atom was lost during imaging, the time of loss, and subsequently the brightness, is described by exponential decay.
\begin{equation}
    \begin{aligned}
PDF_{expDecay}(x) = \begin{cases}
\frac{p^{\frac{x-\mu_0}{d}}\cdot ln(p)}{d\cdot(p-1)} &\text{if $x\in [\mu_0,\mu_1]$}\\
0 &\text{otherwise}
\end{cases}
    \end{aligned}
\end{equation}
for a imaging survival probability p, mean of the peak of unoccupied sites $\mu_0$, mean of the peak of occupied sites $\mu_1$, and a resulting distance of the peaks $d=\mu_1-\mu_0$. Convolving with the Gaussian yields
\begin{equation}
    \begin{aligned}
PDF_{lost}(x) = PDF_{readout} \ast PDF_{expDecay}(x)\\
= \int_{-\infty}^\infty PDF_{expDecay}(t)\cdot \frac{1}{\sigma\cdot\sqrt{2\pi}}\cdot e^{-\frac{1}{2}\cdot (\frac{x-t
}{sigma})^2}dt\\
=\int_{\mu_0}^{\mu_1} \frac{p^{\frac{t-\mu_0}{d}}\cdot ln(p)}{d\cdot(p-1)}\cdot \frac{1}{\sigma\cdot\sqrt{2\pi}}\cdot e^{-\frac{1}{2}\cdot (\frac{x-t
}{sigma})^2}dt\\
= \frac{ln(p)\cdot p^\frac{t-\mu_0}{d}\cdot e^\frac{\sigma^2\cdot ln(p)^2}{2\cdot d^2}\cdot(f(\mu_0)-f(\mu_1))}{2\cdot d\cdot(p-1)}\\
\text{with } f(t) = erf(\frac{x-t}{\sqrt2\cdot\sigma} + \frac{\sigma\cdot ln(p)}{\sqrt2\cdot d})
    \end{aligned}
\end{equation}

In reality, the distribution from exponential decay would have to be convolved with a Poisson distribution before the readout due to shot noise. However, since the variance of the shot noise depends on the number of photons, this is not easily possible. For fitting purposes, convolving only with a Gaussian should not make a huge difference. If more precision is required, one could split the histogram from Fig.~\ref{figures:histogramFittingReal} and conduct two function fittings around the peaks, using different variances for the Gaussian of this distribution of lost atoms. This is not done here.

Using the different theoretical \acp{PDF}, Fig.~\ref{figures:histogramFittingContributions} shows how the total distribution is divided among the individual ones. 
\begin{equation}
    \begin{aligned}
PDF_{total}(x) = a\cdot PDF_{occupied}(x)\\
+ b\cdot PDF_{unoccupied}(x)\\
+ c\cdot PDF_{lost}(x)\\
a + b + c = 1
    \end{aligned}
\end{equation}
where a, b and c represent the fraction of atom locations that are unoccupied, occupied and lost during imaging respectively. For Fig.\ref{figures:histogramFittingContributions}, $a\approx0.393$, $b \approx 0.243$ and $c \approx 0.364$. The imaging survival probability equals $p = \frac{b}{b+c}\approx 0.400$.

Next, the average brightness of imaged atoms is analysed. To begin with, the difference between the x-coordinates of the two peaks may be taken. This can be converted to the average number of photons that were received within the \ac{ROI} during the exposure time for an occupied site. In this example, the first peak is centered around $1.21 \cdot 10^4$ and the second around $2.29 \cdot 10^4$, resulting in a distance of $1.08 \cdot 10^4$. Factoring in the preamp gain of the camera of 4.85, its quantum efficiency at the corresponding wavelength of about 0.86, and the \ac{EM} gain of 300 results in $2.02\cdot 10^2$ received photons during 0.08s of exposure time. Dividing this value by the fractional solid angle of the microscope lens yields the total number of emitted photons. Given the numerical aperture NA, this fractional solid angle $\Omega$ is given by
\begin{equation}
    \begin{aligned}
\Omega = \frac{1 - \sqrt{1-\text{NA}^2}}{2}
    \end{aligned}
\end{equation}

In this project's case, $\text{NA} = 0.65$ and $\Omega \approx 0.120$. The total number of photons emitted per atom during the exposure time would, therefore, be about $1.69 \cdot 10^3$. Per second, that equates to $2.11 \cdot 10^4$ photons per atom.

However, it was not yet considered that not all photons that hit the camera sensor lie within the \ac{ROI}. For diffraction-limited optical systems, it is known that 83.8\% of light is contained within the innermost dark airy ring \cite{Born:1} and that the \ac{ROI} includes at least this first ring. Compensating for the number of photons that may end up outside the \ac{ROI} increases the total number to a maximum of about $2.52 \cdot 10^4$. Also, since aberrations are present in this case, more photons are scattered to outside the \ac{ROI}. Finding the concrete \ac{PSF}, as seen in Fig.~\ref{figures:fittedZernikeFitted}, and calculating the fraction of light within the \ac{ROI} yields a value of 62.3\%. Using this value yields a total number of photons per atom per second of about $3.39\cdot 10^4$. Generating 2000 images using this scattering rate results in a peak separation of $1.17\cdot 10^4$, indicating that overall, the optical response might be slightly less spread out than was configured here.

The brightness in the absence of atoms can also be derived from the pixel \ac{ROI} histogram. By fitting a Poisson distribution convoluted with a Gaussian to the peak of unoccupied atom sites, the average number of spurious charges can be acquired. In this case, this results in an average of 2.62 counts per binned pixel, which equals 0.655 counts or 3.18 photons per physical pixel. By comparing datasets with different exposure times, one can calculate how much of this scales with time and how much is offset. The \ac{CIC} chance, for example, should be similar, regardless of exposure time. To acquire a more precise rate of background illumination, one could also compare image datasets taken with a closed and opened shutter respectively, without loading any atoms.
\subsection{Optical Aberrations}\label{chapters:opticalDefects}
Since the \ac{PSF} represents the response of the optical system to a point-like light source, Zernike polynomials can be fitted to each individual image detail without having to convolute an image. Therefore, only one Fourier transform is used per execution. For this fitting, the Zernike terms with radial degree one to four were used. Piston was ignored because it has no influence on the image. While tilt also does not influence the shape of the optical response, it can be used to allow for precise positioning. Without it, the fitting would always be imprecise, as the fitted image response would always be shifted slightly when compared to the real image. Its value is, however, ignored since it does not represent any real world phenomenon but rather functions as a tool to improve accuracy.

\begin{figure}
  \subfloat[]{\includegraphics[width=0.48\linewidth]{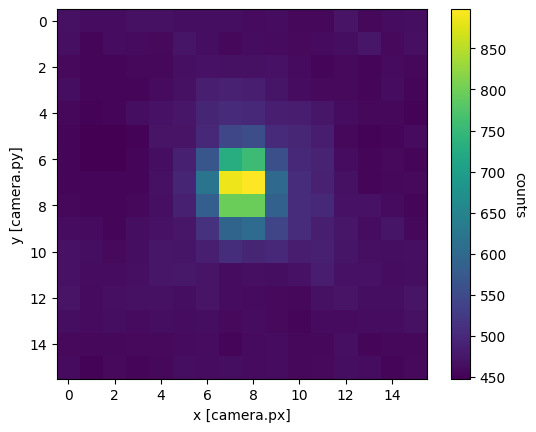}\label{figures:fittedZernikeReal}}
  \hfill
  \subfloat[]{\includegraphics[width=0.48\linewidth]{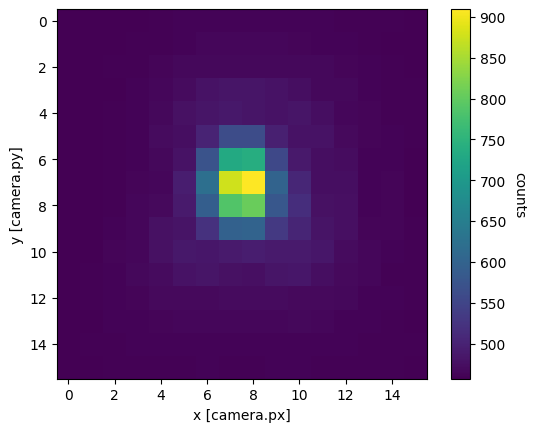}\label{figures:fittedZernikeFitted}}
\caption{(a) Average of 2,864 images of an atom site (b) Visualization of corresponding fitted Zernike polynomials} \label{figures:fittedZernike}
\end{figure}
Fig.~\ref{figures:fittedZernikeReal} shows a small section of the average over 2,864 images with atoms. In Fig.~\ref{figures:fittedZernikeFitted}, one can see a generated image detail with aberrations described by the fitted Zernike polynomials. Table~\ref{table:zernikeCoefficients} shows the corresponding fitted values for the different Zernike coefficients. It is to be noted that blurriness of the optical response is not necessarily caused by actual defocus, but may result from movement of the atoms within their traps. However, there does not appear to be a reason to not describe this via defocus.
\begin{table}
\caption{Zernike coefficients fitted on real data}
\label{table:zernikeCoefficients}
\begin{center}
\begin{tabular}{|p{90pt}|c|c|}
\hline
effect & value & abs. uncertainty \\ 
\hline
defocus (DE) & 0.07232454 & 0.00080439 \\
oblique astigmatism (OA) & 0.00087644 & 0.00106079 \\
vertical astigmatism (VA) & -0.01069755 & 0.00094172 \\
vertical coma (VC) & 0.00280808 & 0.00113738 \\
horizontal coma (HC) & 0.00723265 & 0.00119527 \\
vertical trefoil (VT) & 0.00436401 & 0.00103649 \\
oblique trefoil (OT) & 0.00117688 & 0.00103851 \\
primary spherical (PS) & 0.02449155 & 0.00105728 \\
vertical secondary astigmatism (VSA) & -0.00427388 & 0.00113456 \\
oblique secondary astigmatism (OSA) & -0.00250116 & 0.00109933 \\
vertical quadrafoil (VQ) & -0.00477205 & 0.00135134 \\
oblique quadrafoil (OQ) & -0.00054310 & 0.00134562 \\
\hline
\end{tabular}
\end{center}
\end{table}
\subsection{Camera Characteristics}
In an ideal scenario, all parameters that are specific to the camera should be known from its datasheet. However, in case there is uncertainty for some values or one just wants to confirm their precision, it is possible to fit them to real data as well. For example, the \ac{EM} gain can be fitted to the exponential tail in pixel value histograms. Ideally, one takes images without additional light sources or atoms, so that the overwhelming majority of pixels are hit by no more than one photon.
\begin{figure}
  \includegraphics[width=\linewidth]{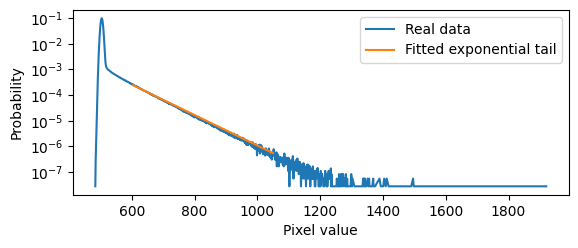}
    \caption{Pixel histogram of image without atoms with fitted exponential tail} 
    \label{figures:tailFitting}
\end{figure}

Fig.~\ref{figures:tailFitting} shows the histogram of individual pixel values and an exponential function that is fitted to it. The fitting function is given by $f(x) = \frac{s}{g}\cdot e^\frac{x}{g}$ for a scale factor s and the \ac{EM} gain $g\approx73.2$. For the images in question, the preamp gain was set to 4.11, resulting in a electron gain of $4.11\cdot g\approx301$. The camera was configured to have a gain of 300.\\
In the same histogram, the left peak could be used to verify the standard deviation of the readout Gaussian distribution since most pixels in this category are not hit by any photon and only differ by readout variations.
\section{Implementation}
We implemented the core algorithms using the methods discussed above in C to allow for maximal performance on modern systems. They are made available in the form of a library via both a C API as well as via Python bindings,
%The interfaces to the simulator library are 
%The algorithm is 
%available both as a C library, as well as a Python package, 
in order to accommodate different users and to allow for both performance and accessibility. Our implementation is freely available on GitHub~\cite{GitHub:1}. 

\subsection{C Library}
The core implementation of the simulation is written in C and has been tested with Microsoft Visual Studio and \ac{GCC}. This C library only depends on FFTW for computing Fourier transforms~\cite{Frigo:2}. For both camera types, a function exists that generates a simulated image given a set of atom sites. Settings that concern the underlying experiment are configured globally.

In the scope of this project, a fast C implementation was deemed necessary. There are several factors that contribute to the overall execution time. First, it scales with the image size. Each pixel samples its own camera noise, leading to a runtime linear to the number of pixels, but the limiting factor is the Fourier transform, which runs in $O(n\cdot log(n))$ for n pixels~\cite{Frigo:1}. Second, the number of photons is important. Some sampling functions take longer for higher numbers of primary photoelectrons. Sampling the \ac{EM} gain, for example, is easily solved for a single primary electron, but becomes more time consuming in case several are present.

\subsection{Python Package}
To make this tool as accessible as possible, we added a Python wrapper that encapsulates the shared library. While the C library offers functions that generate images for the given settings, the Python wrapper offers several classes to the user since readability is more important than speed here. This makes it easy to, for example, preconfigure two different cameras and interchange them between images. The wrapper is available as a Python package from our GitHub, along with detailed lists of the specific classes, attributes, and functions as a Doxygen-based documentation.

Additionally to the actual simulator, we use Python for analyzing data. In particular, we used SciPy for curve fitting~\cite{SciPy:1}, for example when determining the Zernike coefficients.
\section{Evaluation}\label{chapters:evaluation}
Every aspect of the image generation process is subject to stochastic processes. Therefore, we cannot directly compare real images with simulated ones, since we do not aim for them to be exactly the same. They only ought to adhere to the same underlying probability distributions and processes. This section shows to which degree this is actually the case. For some effects like occupancy and positioning, it is sufficient to discuss limitations. Other effects, like the optical response, need to be tested in order to check whether the algorithm accurately produces images that match the configured inputs. All sampling functions have been successfully validated using their theoretical \ac{PDF}.

One property of real images that the simulator is not able to fully reproduce is the continuity of atom coordinates. Since each atom is considered to be point-like prior to the Fourier transform, positioning with subpixel precision is not possible. It is possible to alter the values for tilt in the Zernike coefficients, but this changes positioning for all atom sites equally. In the implementation, this issue is addressed by allowing for further approximation steps, in which the pixels are subdivided for the photon sampling step and combined before the \ac{EM} and readout stage. While this does not fully represent any configuration of atom locations in real data, the combination of discrete subpixel steps and global tilt values should cover the range of pixel alignment of atom sites sufficiently, even if positioning continuously is not possible locally.

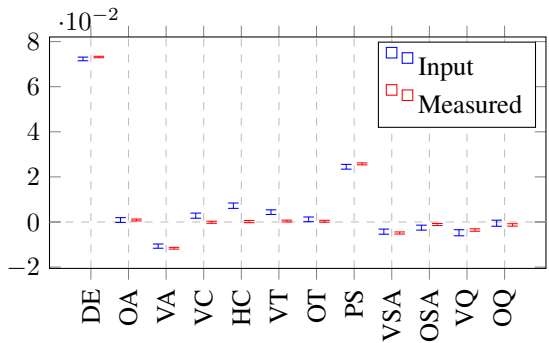
\begin{figure}
\centering
\begin{tikzpicture}
\begin{axis}[ybar=-4pt,symbolic x coords={DE,OA,VA,VC,HC,VT,OT,PS,VSA,OSA,VQ,OQ},xticklabel style={rotate=90},xtick=data,x=0.5cm,no markers,xmajorgrids=true,grid style=dashed,legend cell align={left},legend entries={a, b},scatter/classes={a={mark=square,blue, only marks},b={mark=square,red, only marks}},y=30cm,extra y ticks=0,extra y tick labels=,extra y tick style={grid=major},]
\addplot+[only marks, scatter] plot[error bars/.cd, y dir=both, y explicit]
coordinates{
    (DE, 0.07232454) +- (0.00080439,0.00080439)
    (OA, 0.00087644) +- (0.00106079,0.00106079)
    (VA, -0.01069755) +- (0.00094172,0.00094172)
    (VC, 0.00280808) +- (0.00113738,0.00113738)
    (HC, 0.00723265) +- (0.00119527,0.00119527)
    (VT, 0.00436401) +- (0.00103649,0.00103649)
    (OT, 0.00117688) +- (0.00103851,0.00103851)
    (PS, 0.02449155) +- (0.00105728,0.00105728)
    (VSA, -0.00427388) +- (0.00113456,0.00113456)
    (OSA, -0.00250116) +- (0.00109933,0.00109933)
    (VQ, -0.00477205) +- (0.00135134,0.00135134)
    (OQ, -0.00054310) +- (0.00134562,0.00134562)
};
\addplot+[only marks, scatter] plot[error bars/.cd, y dir=both, y explicit]
coordinates{
    (DE, 0.07317584) +- (0.00032275,0.00032275)
    (OA, 0.00090239) +- (0.00041050,0.00041050)
    (VA, -0.01163870) +- (0.00039345,0.00039345)
    (VC, -0.00012226) +- (0.00046630,0.00046630)
    (HC, 0.00018237) +- (0.00047662,0.00047662)
    (VT, 0.00041168) +- (0.00041415,0.00041415)
    (OT, 0.00033444) +- (0.00041522,0.00041522)
    (PS, 0.02577240) +- (0.00043344,0.00043344)
    (VSA, -0.00487432) +- (0.00046813,0.00046813)
    (OSA, -0.00110784) +- (0.00045264,0.00045264)
    (VQ, -0.00356499) +- (0.00053664,0.00053664)
    (OQ, -0.00130183) +- (0.00053542,0.00053542)
};
\legend{Input,Measured}
\end{axis}
\end{tikzpicture}
\caption{Measured Zernike coefficients from simulated pictures and the coefficients used to generate the images}
\label{figures:measuredZernike}
\end{figure}

To test the correctness of simulated optical responses, the real fitted Zernike coefficients from Section~\ref{chapters:opticalDefects} are used as input for the simulation algorithm. Only one atom site is being used to eliminate crosstalk errors between sites. Generating 2,864 images, the same number as the original dataset, and measuring the Zernike coefficients resulted in the red values in fig.~\ref{figures:measuredZernike}. It can be seen that nearly all values have overlapping uncertainty ranges with the original values from Table~\ref{table:zernikeCoefficients}. Most notable are the differences in the coefficients for coma and vertical trefoil. Nevertheless, there is not a single value that is vastly different from the corresponding original one and the most pronounced defects, such as defocus, astigmatism, and primary spherical, are all reproduced accurately by the simulation. Value uncertainty is improved by a factor of 2 to 2.5 for all coefficients. This is likely due to Zernike polynomials being fixed for the simulation, which might not necessarily be the case for real images. Overall, the comparison provides confidence about the algorithms capability of reproducing the configured optical aberrations.
To check the correlation between photon scattering rate of the atoms and brightness in the final image, we generate 2,000 images with a scattering rate of 30,000 photons per second and all Zernike coefficients set to zero, so that the optical response is diffraction-limited. Following the calculations from section~\ref{chapters:experiment}, one would expect a peak separation between $30000\frac{1}{s}\cdot\frac{0.08s\cdot0.120\cdot0.86\cdot300}{4.85}\approx1.53\cdot10^4$ and ${1.53\cdot10^4\cdot 0.838\approx1.28\cdot 10^4}$. Fitting the peaks yields a separation of $1.43\cdot10^4$, fitting perfectly within the expected range. Compared to the total amount of photons that should hit the camera sensor per atom, about 93.3\% were detected within the \ac{ROI}.

Also, in the simulator, the optical aberrations are the same for all parts of the image. In a real setup, this would not necessarily be the case. For example, it is conceivable that the atom plane may not be perfectly parallel to the focal plane, leading to a gradually changing defocus across the image.

Another possible discrepancy between simulation and real images are the pixel inhomogeneities for \ac{CMOS} cameras. While Section~\ref{chapters:cameraProcesses} goes into great detail about the theory behind all modeled effects, reality looks slightly different.
\begin{figure}
\centering
  \includegraphics[width=0.8\linewidth]{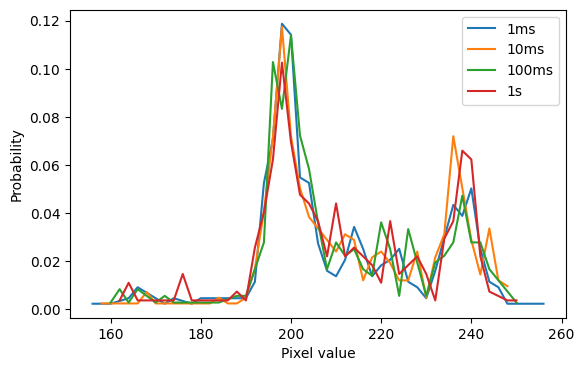}
    \caption{Histogram over all pixel values of a certain physical pixel for exposure times 1ms, 10ms, 100ms and 1s} 
    \label{figures:CMOSTwoPeaks}
\end{figure}

Taking from the same dataset as Fig.~\ref{figures:CMOSClosedShutter} the values from a single physical pixel and creating a histogram shows some strange behaviour. Two separate peaks appear in the probability distribution, as can be seen in Fig.~\ref{figures:CMOSTwoPeaks}. The same effect also appears for different exposure times. Since the shutter is closed, this is not an actual photon distribution, the pixel simply assumes values centered around these two peaks seemingly at random. At the time of writing, it is speculated that this might be due to the fact that there exist two \acp{ADC} that read out each pixel value. We do not, however, understand the whole process behind this and can, therefore, not model it accurately. The \ac{CMOS} simulation created in this project ignores this phenomenon.
\section{Related Work}\label{chapters:relatedWork}
In most work about image reconstruction algorithms, all images used are produced in a real system, which is preferable but requires exclusive access to a quantum experiment. Nevertheless, while the concept of an image simulation is quite new, is not unprecedented. \citeauthor{Rooij:1}~\cite{Rooij:1} show such an image simulation as the base for testing different detection algorithms. However, since their solution mostly focuses on the reconstruction, their image generation is neither developed open-source, nor includes intricacies like optical aberrations. The most important processes like photon distribution and \ac{EM} gain seem to be modeled accurately.

The manufacturer of the \ac{CMOS} camera used in this project, Hamamatsu Photonics K.K., has itself also released an imaging simulation that is available through their website~\cite{Hamamatsu:1}. Their tool allows the user to upload an image and the resulting picture is simulated for different cameras and parameters. However, this tool suffers from similar drawbacks as \citeauthor{Rooij:1}'s version, i.e., it is neither produced open-source, nor does the system allow for precise configuration of optical system and error rates. Additionally, its current web-based form does not make it suitable for use in a tightly coupled feedback loop. It seems to be more of a demonstrator that shows the fundamental differences between the outputs of different camera types.

Considering that there is at least one other group that used a simulator to test deconvolution algorithms, there seems to be a need for a solution that allows for generic parameterization and supports different types of cameras.

While no comprehensive solution for this kind of imaging simulation exists at the moment, many works of research describe the stochastic properties of their images and cameras, which we can use to understand and model the different effects. For example, \citeauthor{Alberti:1}~\cite{Alberti:1} explain different noise sources, mention the probability distribution of electron counts after the \ac{EM} gain, and explain the intricacies of an exemplary optical system. \citeauthor{Madjarov:1}~\cite{Madjarov:1} describes the scattering of photons for cold atom fluorescence imaging and gives a measure for the rate of photons that one can expect per imaged atom. These sources and more were considered for the simulator's implementation.
\section{Conclusions}
In this project we developed a tool that can produce sample images of neutral atom quantum computer setups for exemplary purposes or to calibrate and train reconstruction algorithms. While small inconsistencies exist, the final image simulation is a success. The algorithm is fit for any usage that is intended purely for inspection with the human eye since any discrepancies would be impossible to discern that way.

However, it must be understood that in order to serve as a tool in developing deconvolution algorithms, this is not necessarily sufficient. Any user should take the imperfections discussed in Section~\ref{chapters:evaluation} into account and consider them in their particular use case.
%the way that the simulator can be used in their case. 
However, there should hardly be any case for which any of these differences represents a problem. If one wants to confront a deconvolution algorithm with different optical responses, even though the simulated aberrations can only be configured globally, one could still generate different images across ranges of Zernike coefficient values. It should not make a difference whether the whole range of aberrations appear within a single image or are represented by changing parameters across a set of images. Apart from the unexplained effects of \ac{CMOS} cameras, all camera noise has been modeled accurately and the optical system can be tuned quite finely.

%If a user thinks that an effect may not be modeled accurately enough, then the algorithm can at least function as a base to allow them to expand on the implementation of the problem area themselves.

For the main use cause, the ability to automatically generate images with an associated ground truth, i.e., automatically labeled data, the presented techniques are sufficient.
%The most important improvement that is introduced by the simulator is undoubtedly the possibility of acquiring labeled data. 
Being able to train or verify a reconstruction algorithm's output opens up many possibilities and the ability of being able to make meaningful claims of detection fidelities will be critical in developing efficient and fast neutral atom quantum computers.

\section*{Acknowledgments}
We thank the Strontium Rydberg Lab at \ac{MPQ} for providing all image data for this publication and especially its members Maximilian Ammenwerth and Flavien Gyger for discussions about the results. We also thank the \ac{MQV} team at \ac{MPQ} for many fruitful discussions.

% Generated by IEEEtranN.bst, version: 1.13 (2008/09/30)

\end{document}